\documentclass{article}
\usepackage{spconf,amsmath,graphicx}
\usepackage{amssymb}
\usepackage{bm}

\title{UniSpeech at scale: An Empirical Study of Pre-training Method on large-scale speech recognition dataset}
\name{Chengyi Wang\sthanks{The work was done as an intern at Microsoft}, Yu Wu, Shujie Liu, Jinyu Li, Yao Qian, Kenichi Kumatani and Furu Wei}
\address{ Microsoft}
\begin{document}
%
\maketitle
\begin{abstract}
Recently, there has been a vast interest in self-supervised learning (SSL) where the model is pre-trained on large scale unlabeled data and then fine-tuned on a small labeled dataset. The common wisdom is that  SSL helps resource-limited tasks in which only a limited amount of labeled data is available. The benefit of SSL keeps diminishing when the labeled training data amount increases. To our best knowledge, at most a few thousand hours of labeled data was used in the study of SSL. In contrast, the industry usually uses tens of thousands of hours of labeled data to build high-accuracy speech recognition (ASR) systems for resource-rich languages. In this study, we take the challenge to investigate whether and how SSL can improve the ASR accuracy of a state-of-the-art production-scale Transformer-Transducer model, which was built with 65 thousand hours of anonymized labeled EN-US data.
\end{abstract}
\section{Introduction}
\label{sec:intro}
In the past decade, the speech recognition field has made huge progress owing to the  deep learning techniques \cite{DNN4ASR-hinton2012}. However, the development cost of automatic speech recognition (ASR) system is still high due to the requirement of large-scale labeled data. To tackle the problem, researchers investigate different semi-supervised and unsupervised approaches, such as teacher-student learning \cite{chebotar2016distilling, li2018developing, movsner2019improving}, self-supervised learning (SSL) \cite{DBLP:journals/taslp/ChenHLWS19,DBLP:journals/taslp/ChorowskiWBO19,oord2018representation, DBLP:conf/interspeech/SchneiderBCA19, DBLP:conf/iclr/BaevskiSA20, DBLP:conf/nips/BaevskiZMA20} and self-training \cite{xu2021self, zhang2020pushing}. 

Recently,  SSL has drawn lots of attention and achieved great success in the low-resource ASR setting. The training process usually is to first pre-train the model with large amount of unlabeled data and then fine-tune the model with relatively small amount of labeled data. Wav2vec 2.0 \cite{DBLP:conf/nips/BaevskiZMA20} shows that pre-training with 60k hours unlabeled audio data can significantly boost the accuracy on the 960 hours Librispeech benchmark. Furthermore,
\cite{baevski2021unsupervised} demonstrates that the  SSL performs very well in a pure unsupervised setting. In addition, a bunch of work \cite{DBLP:conf/icassp/RiviereJMD20, DBLP:journals/corr/abs-2006-13979,wang2021unispeech} yields impressive results in different low-resource ASR settings, such as multilingual ASR and domain shift ASR.

The works of SSL focus on how to improve ASR performance when only limited transcribed data is available and the gain of SSL keeps diminishing when the labeled training data amount increases \cite{DBLP:conf/nips/BaevskiZMA20}. To our best knowledge, at most a few thousand hours of labeled data was used in the study of SSL \cite{chan2021speechstew}.  In contrast, the industry usually uses tens of thousands of hours of labeled data to build high-accuracy ASR systems for resource-rich languages  \cite{sainath2020streaming, Li2020Developing}. Besides, in the previous work, both the pre-training and fine-tuning are in offline mode, where the whole utterance is required to generate the transcription. However, the online ASR system is crucial for many industry scenarios. In this paper, we focus on the questions that whether the  SSL method can improve ASR performance when large-scale labeled data is available and whether the previous SSL method can help streaming ASR system. To explore SSL in a large-scale setting, we conduct experiments with Microsoft internal anonymized datasets, containing 65 thousand (K) hours supervised data and  155K unlabeled speech data. 
To prepare unlabeled data, we elaborately collect diverse audios, such as noisy background audios, accent speaker speech, as well as 8K sampling rate audios. The 65K labeled data are mostly native speaker speech with cleaner background and 16K sampling rate, which will be referred as in-domain data in this paper. For pre-training, we mix up the labeled and unlabeled data, resulting in a speech corpus with 220K hours data. To test the performance of SSL on online system, we fine-tune the model in a streaming fashion and test it on both in-domain and out-domain test sets. 
As far as we know, this paper is the first one to explore the effectiveness of pre-training method when very large-scale supervised data is available. It will help industry and academia to better understand the contribution of speech pre-training method in  model building with huge supervised and unsupervised data.

We employ Transformer Transducer model \cite{yeh2019transformer,zhang2020transformer, chen2021developing} as the backbone model structure. In the pre-training stage, we conduct the multi-task learning proposed in our previous work UniSpeech \cite{wang2021unispeech}, where transducer loss and contrastive loss are combined together. To stablize and accelerate the training process, we follow \cite{zhang2020pushing} to replace the quantizer component with a linear module. With the pre-trained Transformer encoder, we learn a Transformer Transducer in a streaming fashion in the fine-tuning stage. Experiment results show that
UniSpeech can largely improve the performance on out-domain testsets, while the performance improvement on in-domain testsets is relatively small. Furthermore, we find the multi-task training performs better than contrastive learning on this large-scale dataset.

\section{Related Work}
 \begin{figure*}[t]
    \centering
    \includegraphics[width=1 \textwidth]{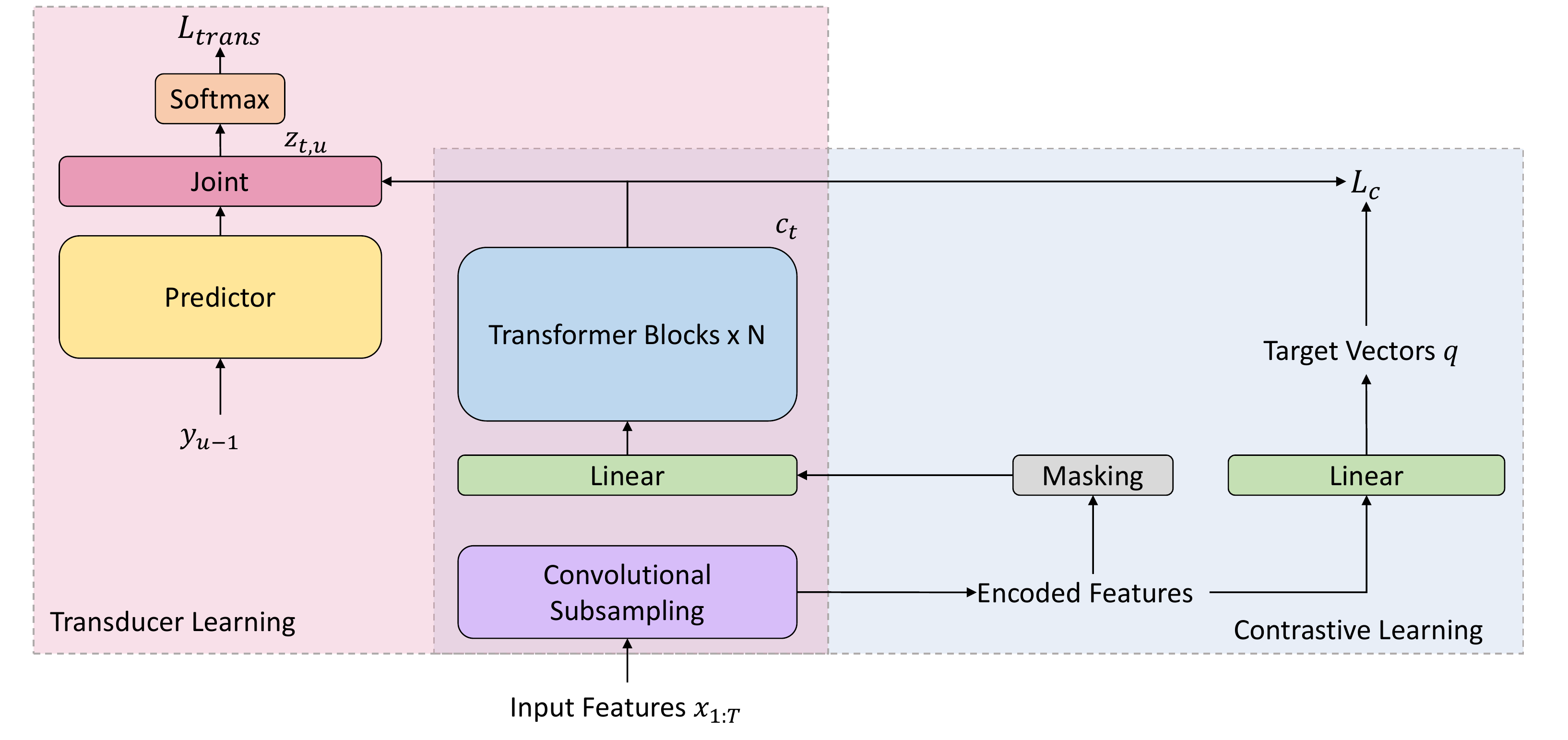}
  \caption{The UniSpeech architecture. The right part with a blue background is contrastive learning. The left part with a red background is transducer learning. The encoder is used for both.}
  \label{model}
\end{figure*}

Recently, there is a large body of work on  SSL in ASR, which can be summarized into two lines. The first line is generative pre-training, which reconstructs audio signals or features.  In this research line, full reconstruction with autoencoders\cite{DBLP:journals/taslp/ChenHLWS19,DBLP:journals/taslp/ChorowskiWBO19}, future reconstruction with autoregressive predictive coding (APC) \cite{DBLP:conf/interspeech/ChungHTG19} and masked reconstruction \cite{DBLP:conf/icassp/LiuYCHL20,DBLP:conf/icassp/LingLSK20,ling2020bertphone} has been investigated. Another research line is discriminative pre-training \cite{oord2018representation,DBLP:conf/interspeech/SchneiderBCA19,DBLP:conf/iclr/BaevskiSA20}, which is highly related to our paper. In this line, contrastive predictive coding (CPC) \cite{oord2018representation} uses an autoregressive model to classify future frames from negative examples. Wav2vec \cite{DBLP:conf/interspeech/SchneiderBCA19} evaluates the effectiveness of contrastive learning on the ASR task.  \cite{DBLP:conf/emnlp/KawakamiWDBO20} and \cite{DBLP:conf/icassp/RiviereJMD20} show bi-directional and modified CPC transfers well across domains and languages.  Vq-wav2vec \cite{DBLP:conf/iclr/BaevskiSA20} uses a vector quantization module to learn discrete representations. \cite{DBLP:conf/nips/BaevskiZMA20} proposes wav2vec 2.0 which masks the speech input in the latent space and solves a contrastive task defined over contextual representations in the masked region and a quantization of the latent representations. It shows that pre-training with 60K unlabeled audio data can significantly improve the accuracy on the Librispeech benchmark. Its effectiveness has also been verified on multilingual setting \cite{DBLP:journals/corr/abs-2006-13979} and when combined with self-training \cite{xu2021self, zhang2020pushing}.  Recently, \cite{baevski2021unsupervised} proposes wav2vec-U and proves that unsupervised learning can performs well on ASR task. Our previous work UniSpeech \cite{wang2021unispeech} explores pre-training on low-resource languages and domain transfer tasks. It learns speech representations in a multitask manner with both labeled and unlabeled data. Most of the previous works verify the effectiveness of  SSL on low-resource scenario where the labeled data is limited. And most of them train an offline ASR system. It is unclear whether the method works on high-resource setting and on streaming setting. In this work, we investigate UniSpeech on a very challenging setting in which 65K hours labeled data is used to build the streaming supervised training baseline.

 \begin{figure}[t]
    \centering
    \includegraphics[width=0.5 \textwidth]{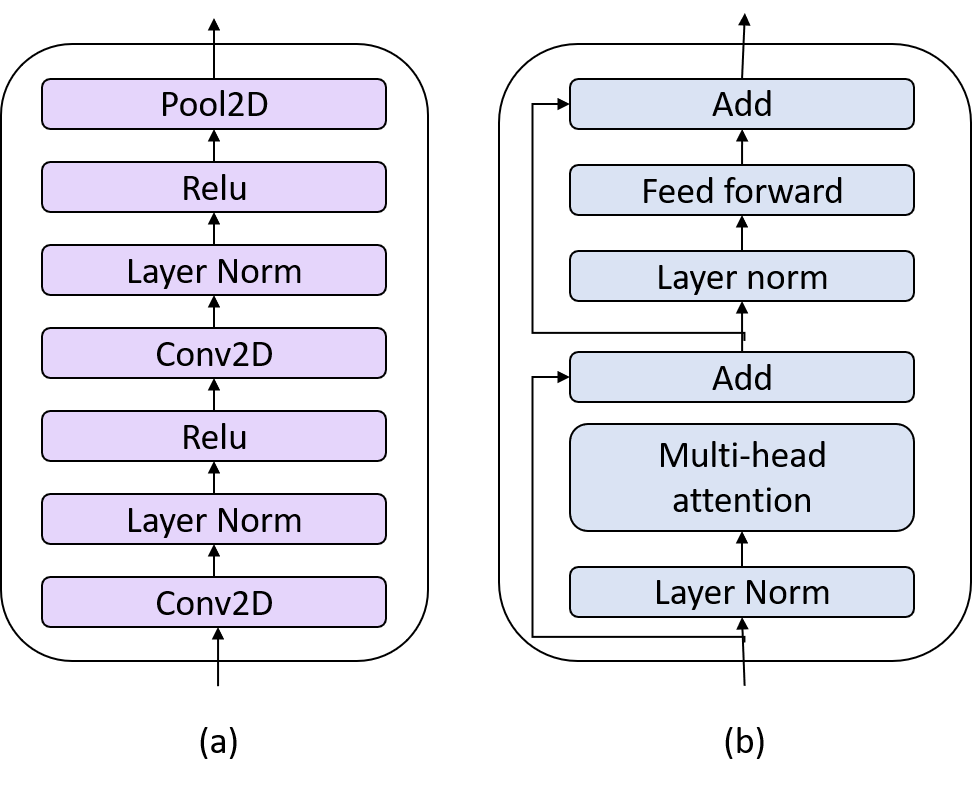}
  \caption{(a) Convolutional feature encoder block. (b) Transformer block. }
  \label{module}
\end{figure}

\section{UniSpeech}

Figure \ref{model} shows the architecture of UniSpeech, with Transformer Transducer as its backbone ASR model which contains a Transformer-based encoder, an LSTM-based prediction network and a joint network.   Given an input speech sequence of length $T$, $\bm{x} = (x_1, ..., x_T)$, the transducer model tries to predict the target label sequence $\bm{y}=(y_1, ..., y_U)$. The encoder is also used for contrastive learning. We will first introduce the details of model structure, and then show pre-training and fine-tuning paradigm. 

\subsection{Model Structure}

The encoder consists of a convolutional feature encoder and a context network. It converts the audio input $\bm{x}$ into a high-level repraesentation $\bm{c}$. The feature encoder has two convolutional blocks. As shown in Figure \ref{module} (a), each block contains two 2D convolutional layers followed by layer normalization and RELU activation layers. All the convolution layers have kernel size (3, 3) and stride (1, 1). For the first block, the output channel sizes are 64 for the two 2D convolution. For the second block, the output channel sizes are 128. There is a 2D max-pooling layer at the top of each block. The downsampling rate is 2 for the first block and 4 for the second block, resulting in an 8x reduction in the feature sequence length.  After downsampling, the encoded features are transformed by a linear layer and then fed into the Transformer context network. We use 18 stacked Transformer blocks in our work. Each block contains two sub-layers, a multi-headed attention layer and a feed-forward layer. The multi-head attention projects the input to query $q$, key $k$, and value $v$ and compute self-attention separately for each head. Then the weight-averaged $v$ from all heads are concatenated and passed to the feed-forward sublayer. Our model has two modifications with the original Transformer block \cite{vaswani2017attention}: First, we apply layer normalization before the multi-head attention and feedforward layers instead of after (i.e. $x+\text{AttentionLayer}(\text{LayerNorm}(x)))$. This has been unanimously adopted by many Transformer implementations leading to more effective
training \cite{DBLP:conf/iclr/BaevskiA19, DBLP:conf/icml/XiongYHZZXZLWL20}. Secondly, we use relative attention as used in \cite{DBLP:conf/naacl/ShawUV18} instead of sinusoidal positional embeddings as it was found to yield better performance. The motivation is the offset between two frames should be considered in the attention weight calculation and the offset is modeled by the relative position embedding. For efficiency, we use a simple but effective relative position embedding, which is formulated as 
\begin{equation}
    \alpha_{t, \tau} = \text{Softmax}(q^T_t(k_\tau + p_{t, \tau})v_{\tau}
\end{equation}
where $\alpha_{t, \tau}$ is the self-attention weights between position $t$, $\tau$ and $p_{t, \tau}$ is the relative position embedding obtained from a lookup table. We use model dimension of 512, feed-forward inner dimension of 2048 and 8 attention heads.

The prediction network acts like a language model. It produces a representation $h_u$ conditioning on the previous non-blank target $y_{u-1}$ predicted by the model. Our prediction network has 2 stacked blocks, each of which contains a uni-directional LSTM, a linear projection layer and a linear normalization layer. The LSTM has a memory cell size of 1024, and the projection layer reduces the dimension to 640. 

The joint network is a feed-forward network that combines the encoder output $c_t$ and the prediction network output $h_u$ as 
\begin{equation}
    z_{t,u} = f^{joint}(c_t, h_u)
\end{equation}

\subsection{Multitask Pre-training}
Supposing we have an unlabeled datasets $\mathbb{M}$ and a labeled dataset $\mathbb{L}$,
the pre-training objectives can be split into three parts: 1) Contrastive loss $\mathcal{L}_{c}$ on dataset $\mathbb{L}$, 2) Contrastive loss on dataset $\mathbb{M}$ and 3) Transducer loss $\mathcal{L}_{trans}$ on dataset $\mathbb{L}$.  We will first introduce how to compute contrastive loss and transducer loss respectively and then show the loss combination method.

\noindent \textbf{Contrastive Learning.}
Given the speech input $\bm{x}$, we can obtain the downsampled feature representation from the convolutional feature encoder. In contrastive learning, we mask some frames and fed the masked features into the Transformer. We use the same mask strategy as in wav2vec2.0 that randomly sample start indices with probability $p$ and masks the consecutive ten time steps. Instead of using a quantizer output as the contrastive targets as in Wav2vec2.0 and UniSpeech, we follow \cite{zhang2020pushing} to replace the quantization module with a linear layer. The targets are denotes as $\bm{q}$. For each encoder output $c_t$ centered over masked time step $t$, the model needs to identify the true quantized latent speech representation $q_t$ in a set of $K+1$ quantized candidates $\rm{Q}_t$. The $K$ distractors are uniformly sampled from the other timesteps in the same utterance. This frees up the model from using its capacity to represent speaker-dependent information and instead focuses on content information.  The loss is defined as 
\begin{equation}\small
    \mathcal{L}_{c} = - \sum_{t\in M_T}\text{log}\frac{\text{exp}(sim(c_t, q_t))}{\sum_{\tilde{q} \sim \rm{Q}_t}\text{exp}(sim(c_t, \tilde{q}))}
\end{equation} where we use cosine similarity $sim(a, b) =\frac{a^Tb}{\| a \| \| b \|} $. $M_T$ represent all  masked positions in an utterance.

\noindent \textbf{Transducer Learning.}
We extend the UniSpeech by replacing the CTC loss \cite{graves2006connectionist} with transducer loss \cite{Graves-RNNSeqTransduction} here as it improves upon CTC by making the output symbol distribution dependent on the previous output history. 

With joint network output $z_{t,u}$, we obtain $h_{t,u}$ with a linear transform and the posterior for each output token $k$ is 
\begin{equation}
 P(k|\bm{x}_1^{t}, \bm{y}_1^{u-1}) = \text{softmax}(h_{t, u}^k)
\end{equation}
The transducer model defines a conditional distribution $P(\bm{\hat{y}}|\bm{x})$ over all the possible alignments, where \begin{equation}
    \bm{\hat{y}} = \{({\hat{y}}_1, t_1), ... (\hat{y}_{T+U}, t_{T+U})\}
\end{equation} is a sequence of $(\hat{y}_i, t_i)$ pairs. $(\hat{y}_i, t_i)$ represents an alignment between output $\hat{y}_i$ and the encoded feature at time $t_i$. Each label $\hat{y}_i$ can optionally be blank label. Removing the blank labels we can get the actual output label sequence $\bm{y}$ with length $U$. 

The final transducer loss is defined as
\begin{equation}
\begin{aligned}
    \mathcal{L}_{trans} &= -\text{ln}P(\bm{y}|\bm{x}) \\
     & =-\text{ln} \sum_{\bm{\hat{y}}\in H(\bm{y}, T)}P(\bm{\hat{y}}|\bm{x}) \\
     & =-\text{ln} \sum_{\bm{\hat{y}}\in H(\bm{y}, T)} \prod_{i=1}^{T+U}P(\hat{y}_i|\bm{x}_1^{t_i}, \bm{\hat{y}}_1^{u_{i-1}})
     \end{aligned}
\end{equation}
where $H(\bm{y}, T) $ corresponds to the set of all possible paths.

\noindent \textbf{Loss Combination.} During pre-training, we sample batches from $\mathbb{L}$ and $\mathbb{M}$ alternatively. For batch from labeled dataset $\mathbb{L}$, we compute both the contrastive loss and the transducer loss. For batch from unlabeled dataset $\mathbb{M}$, we only compute the contrastive loss. The final pre-training loss can be defined as
\begin{equation} \label{multi_task}
\small
    \mathcal{L} =  \sum_{(\bm{x},\bm{y}) \in \mathbb{L}}(\alpha \mathcal{L}_{trans} + (1-\alpha) \mathcal{L}_{c}) + \sum_{(\bm{x}) \in \mathbb{M}} \mathcal{L}_{c} 
\end{equation} where $\alpha$ is the weight for loss combination on dataset $\mathbb{L}$.
 \begin{figure}
    \centering
    \includegraphics[width=0.35 \textwidth]{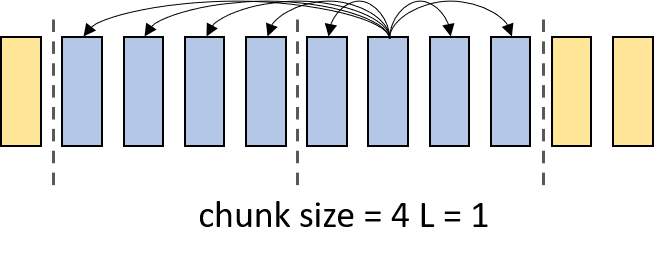}
  \caption{Chunk-wise self-attention.}
  \label{streaming}
\end{figure}
\subsection{Fine-tuning}

After pre-training, we fine-tune the model with dataset $\mathbb{L}$ and the transducer loss. The model is trained in a streaming mode that both the left and the right context for self-attention is limited. The attention mechanism provides a flexible way to control the context that the model uses. The attention mask can be applied on the attention weight matrix $\alpha_{t, \tau}$ to determine the range of receptive field. The attention mask is created with the following rules: First, the input acoustic sequence is segmented into chunks with a specified chunk size, and there is no overlap between chunks. The frames in the same chunk can see each other. Each frame can see the frames in the left several chunks within a left chunk window. And the frame cannot attend to other frames in the consecutive chunks. In Figure \ref{streaming}, we show the masked self-attention when chunk size is 4 and left chunk window size is 1, which means the previous one chunk is available for current frame. This method makes the left reception field grow linearly with the model becoming deeper, while the chunk boundary will strictly restrict the right reception field. In our experiment, we set the chunk size to 4 and the left window size is 18. As each frame corresponds to 80 ms audio signals, the latency for self-attention is 320 ms.

\section{Experimental Setup}

\subsection{Datasets and Domains}

The supervised training setup is the same as the setup in our previous study \cite{Li2020Developing},  with 65 thousand (K) hours of transcribed Microsoft data. The  sampling rate of all the data is 16K HZ. For the   SSL, we further add 155K hours unlabeled data, resulting 220K hours data in total when combining both labeled and unlabeled data. All the unlabeled data come from new domains that are not covered in the supervised training setup. The sampling rate of most of these unlabeled data is still 16K HZ , but it is 8K HZ for some data of them. The feature is 80-dimension log Mel filter bank for every 10 milliseconds (ms) speech. Utterances longer than 30 seconds are discarded during training. We use the method in \cite{li2012improving} to convert the Mel filter bank feature of 8K sampled data into 16K sampled feature by zeroing the high frequency bank energy. The output targets are 4K word-piece units.

We have several test sets: The first test set covers 13 application scenarios such as Cortana and far-field speech, with a total of 1.8 million (M) words from 250K utterances. This test set has the same domain as the supervised training dataset. We denote it as Task A and report the word error rate (WER) averaged over these 13 test scenarios. Task B is a dictation task  with 5.8K utterances and 63K words. This test set is also highly related to the supervised training set. Other test sets come from new domains. Task C is a conversation task between two people, sampled at both 16K and 8K HZ frequency, each of which has 400 utterances with 46K words.  We also have several test sets from accent speech. Task D is an accented English task, covering around 20 kinds of accents with a total 1.9M words from 175K utterances.There are also test sets from four English speaking countries, England(EN-GB), Australia (EN-AU), Canada (EN-CA) and New Zealand (EN-NZ), with 740k, 260K, 250K and 26K words from 58K, 50k, 25K and 1K utterances, respectively. Except for Task C, all the other test sets have sample rate 16K HZ. The unlabeled training data is a superset of all these tasks, with a larger amount of additional data covering all kinds of domains we can collect, with both 16K and 8K sampling rate speech. 
All the training and test data are anonymized data with personally identifiable information removed. 


\subsection{Implementation}
\noindent \textbf{Pre-training} We implement models with our in-house transducer codebase. For contrastive pre-training, we sample $p=0.065$ for all time-steps to be starting indices and mask the subsequent 10 timesteps. The number of negative samples are set to 100. We set loss weight to 0.5 in our experiment. The model is pre-trained on 32 V100 GPUs. We use different batch size for different datasets as transducer loss computation requires more memory usage. For labeled data, the batch size is set to 24k frames per GPU. For unlabeled data, the batch size is set to 48k frames. We optimize the model with Adam, warming up the learning rate for 25k steps and then linearly decay it to a total 420k steps. The learning rate scheduler is:
\begin{equation}
    \text{lrate} = k \cdot d_{model}^{-0.5} \cdot \min (n^{-0.5}, n \cdot warmup^{-1.5})
\end{equation}
where $k$ is the learning rate factor, $d_{model}$ is the attention dimension of Transformer block and $n$ is the current step. In pre-training, we set $k$ as 5 and $warmup$ as 25000.

\noindent \textbf{Fine-tuning}
During finetuning, we use the same optimizer and set the learning rate scheduler factor as 6.  The model is trained on 16 V100 GPUs. We group batches by the production of input length and output length and set the maximum batch size as 900k.
\begin{table*}[t]
\begin{center}

\begin{tabular}{l|c|c|c} 
\hline \hline
Test sets                          & Supervised Training   &  UniSpeech &  w/o multitask pre-training   \\ \hline
Task A  &   8.33   &  8.16  & 8.41  \\
Task B  &  15.20 & 15.15   & 15.31 \\ \hline
Task C-16k  &  19.36  & 17.76 & 19.94 \\ 
Task C-8k  &  27.69  & 23.49 & 27.96 \\ \hline
Task D  &   16.18  & 15.52  & 16.61  \\
EN-GB   &   17.90   & 17.73  & 19.11 \\
EN-AU   &   13.50  &  12.95  &13.62 \\
EN-CA  &  12.69  & 12.69 &  13.58  \\
EN-NZ   &   20.03   &  18.59 & 21.15 \\ \hline
Overall  &  13.46   &  13.03 &  13.84 \\ \hline \hline

\end{tabular}\caption{A comparision with training  from scratch. The w/o multitask setting is very similar to the original wav2vec 2.0, but quantization component is removed and fbank is used as the raw feature.  \label{result}}

\end{center}

\end{table*}
\section{Results}

We compare our UniSpeech method with supervised training baseline. The baseline model is trained with only 65K labeled data using transducer loss. The training setup is the same as the fine-tuning process in Section 4.2. The results are shown in Table \ref{result}.

\subsection{In-domain Evaluation}
We first evaluate the model in in-domain scenario. On Task A, UniSpeech and baseline model achieve 8.16\% and 8.33\% WER score. On Task B, UniSpeech and baseline get the 15.15\% and 15.2\% WER respectively. It can be seen that our method is slightly better than supervised training from scratch, but the gain is limited. This result is expected as the test sets are highly represented in the supervised data set. The results show that when the large scale labeled data is available, adding more unsupervised out-of-domain data has limited influence on the in-domain test sets. 

\subsection{Cross-domain Evaluation}
Then we evaluate the models on out-of-domain test sets. On Task C, we test the model on 8K set and 16K set separately.
For the 16K set, UniSpeech obtains 17.76\% WER while the  baseline is 19.36\%. UniSpeech outperforms baseline by  8.26\% relatively. On 8K test set, we can observe a larger improvements where the WER score of UniSpeech and supervised baseline is 23.39\% and 27.69\% respectively, with a relative gain of 15.17\%. The larger gain on the 8k test set is because the unlabeled data set has a large amount of 8k sampled data while the original labeled set doesn't.   This indicates our model can well utilize both supervised and unsupervised data and it learns robust speech representation which can be transferred across different domains.

\subsection{Accent Evaluation}
The accent test sets can also be seen as out-of-domain sets. On Task D, UniSpeech has a relative gain of 4.08\% compared to the baseline. On other accent datasets from English speaking countries, UniSpeech achieves 0.95\%, 4.07\%, 0\% and 7.2\% gain respectively. The conclusion is consistent with cross-domain setting but the gain on accented setting is smaller. It is no surprise that EN-CA doesn't have any gain since EN-CA is very similar to EN-US in the supervised training data.

\subsection{Ablation Study}
 \begin{figure}
    \centering
    \includegraphics[width=0.5 \textwidth]{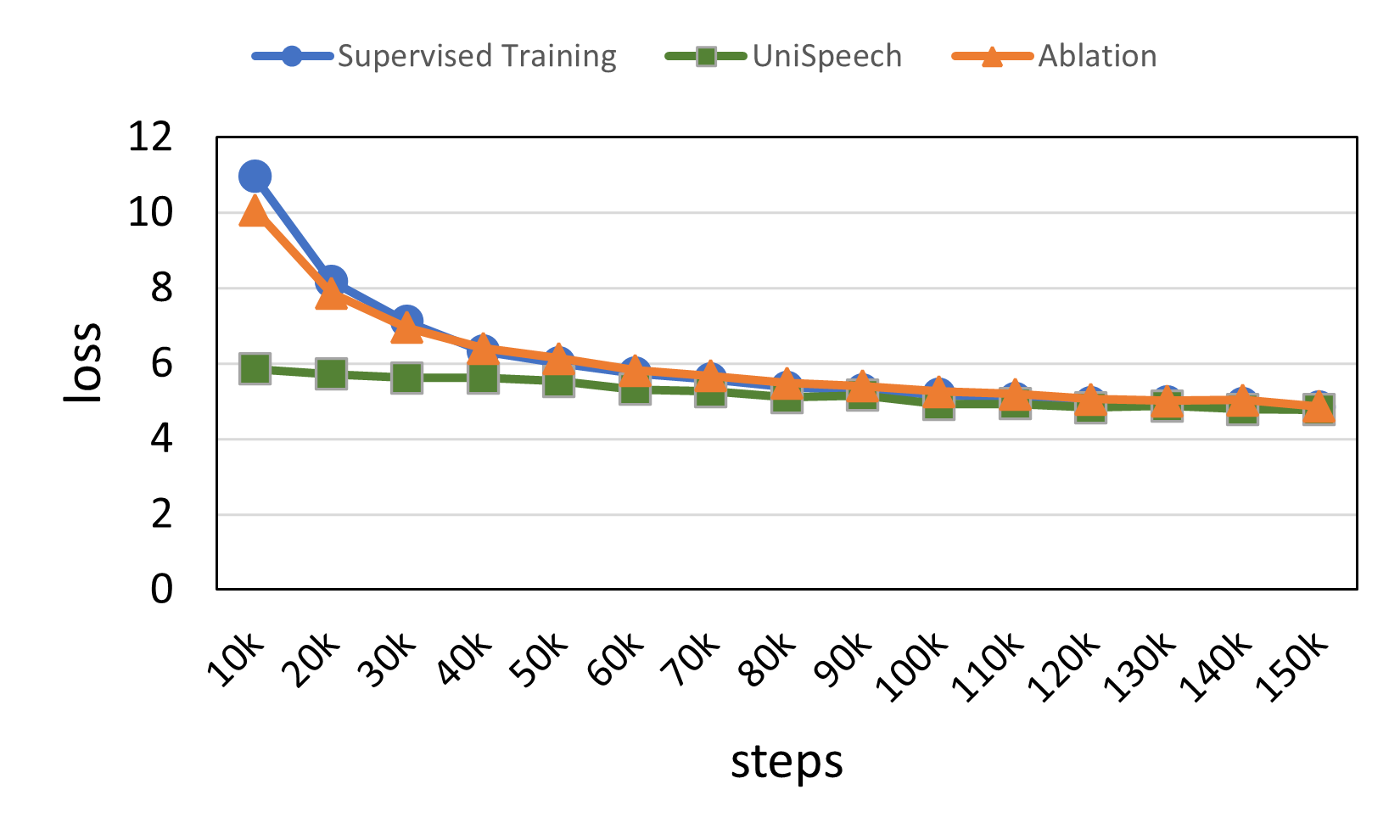}
  \caption{Validation loss curve.}
  \label{loss}
\end{figure}
To further evaluate the effectiveness of multitask pre-training paradigm in our pre-training process, we design an ablation experiment. In this experiment, we pre-train the model towards only contrastive loss with the 220K speech corpus. In other words, we set the loss weight $\alpha$ in Equation \ref{multi_task} to 0.  The w/o multitask setting is very similar to the original wav2vec 2.0 except that the quantization module is replaced with a linear layer and we use fbank as input features. We have conducted an offline experiment to compare our model design with the original wav2vec2.0. We pretrain two models on 960h Librispeech data and then finetune on the 100h clean subset with fairseq \footnote{{https://github.com/pytorch/fairseq}}. Results show that our design achieves the same result on test-clean and less than 10$\%$ degradation on test-other. We find original wav2vec 2.0 is hard to tune on large-scale dataset, so we change to the simple version. 

From this table, we can see that removing multitask pre-training gets comparable results on in-domain Task A and Task B. This further proves that the additional out-of-domain unlabeled data has little effect on high-resource ASR tasks. While on out-of-domain tasks and accented tasks, the performance drops significantly. On Task C, the performance drops relatively 12.2\% and 19.1\%  on the 16K HZ test set and 8K HZ test set respectively. On accented English Task D, the WER drops about 7.0\% relatively. The results on other accented tasks are consistent. This indicates that the gain of UniSpeech is mostly come from the multitask learning manner. When comparing the ablation experiment results with baseline results, we are surprised to find that pre-training only on unlabeled data gets even worse results. One possible explanation is that the pre-training model can see the whole utterance to perform contrastive learning while the fine-tuned model has limited context. It is hard to provide a good start point with only contrastive loss, especially when a large amount of supervised data is available. We further compare the validation transducer loss during fine-tuning stage. The validation set has the same domain as the supervised training data. We draw the  loss curve before 150K fine-tuning steps in Figure \ref{loss}.  We can see that the ablation experiment is slightly better only in the beginning 30K steps. After that, there's no difference between the two model. As the unsupervised pre-training cannot learn ASR specific representations, more pre-training steps may be required to get a better fine-tuning start point. While for UniSpeech, the validation loss is much lower as the prediction network and joint network have been updated during pre-training.


\section{Conclusion}
In this paper, we investigate the effectiveness of pre-training method when large-scale labeled data is available. The pre-training data contains 220K hours audio while the fine-tuning data contains 65K hours labeled speech.  We extend our UniSpeech that uses multitask learning with contrastive loss and transducer loss in the pre-training stage. And then we fine-tune the transducer model on labeled data. Experiments show that UniSpeech can largely improve out-of-domain performance, and the multi-task training performs better than contrastive pre-training on large-scale setting.
\bibliographystyle{IEEEbib}
\bibliography{refs}

\begin{thebibliography}{10}

\bibitem{DNN4ASR-hinton2012}
Geoffrey Hinton, Li~Deng, Dong Yu, George~E Dahl, Abdel-rahman Mohamed, Navdeep
  Jaitly, Andrew Senior, Vincent Vanhoucke, Patrick Nguyen, Tara~N Sainath,
  et~al.,
\newblock ``Deep neural networks for acoustic modeling in speech recognition:
  The shared views of four research groups,''
\newblock {\em IEEE Signal processing magazine}, vol. 29, no. 6, pp. 82--97,
  2012.

\bibitem{chebotar2016distilling}
Yevgen Chebotar and Austin Waters,
\newblock ``Distilling knowledge from ensembles of neural networks for speech
  recognition.,''
\newblock in {\em Interspeech}, 2016, pp. 3439--3443.

\bibitem{li2018developing}
Jinyu Li, Rui Zhao, Zhuo Chen, Changliang Liu, Xiong Xiao, Guoli Ye, and Yifan
  Gong,
\newblock ``Developing far-field speaker system via teacher-student learning,''
\newblock in {\em Proc. ICASSP}, 2018, pp. 5699--5703.

\bibitem{movsner2019improving}
Ladislav Mo{\v{s}}ner, Minhua Wu, Anirudh Raju, Sree Hari~Krishnan
  Parthasarathi, Kenichi Kumatani, Shiva Sundaram, Roland Maas, and Bj{\"o}rn
  Hoffmeister,
\newblock ``Improving noise robustness of automatic speech recognition via
  parallel data and teacher-student learning,''
\newblock in {\em ICASSP 2019-2019 IEEE International Conference on Acoustics,
  Speech and Signal Processing (ICASSP)}. IEEE, 2019, pp. 6475--6479.

\bibitem{DBLP:journals/taslp/ChenHLWS19}
Yi{-}Chen Chen, Sung{-}Feng Huang, Hung{-}yi Lee, Yu{-}Hsuan Wang, and
  Chia{-}Hao Shen,
\newblock ``Audio word2vec: Sequence-to-sequence autoencoding for unsupervised
  learning of audio segmentation and representation,''
\newblock {\em {IEEE} {ACM} Trans. Audio Speech Lang. Process.}, vol. 27, no.
  9, pp. 1481--1493, 2019.

\bibitem{DBLP:journals/taslp/ChorowskiWBO19}
Jan Chorowski, Ron~J. Weiss, Samy Bengio, and A{\"{a}}ron van~den Oord,
\newblock ``Unsupervised speech representation learning using wavenet
  autoencoders,''
\newblock {\em {IEEE} {ACM} Trans. Audio Speech Lang. Process.}, vol. 27, no.
  12, pp. 2041--2053, 2019.

\bibitem{oord2018representation}
Aaron van~den Oord, Yazhe Li, and Oriol Vinyals,
\newblock ``Representation learning with contrastive predictive coding,''
\newblock {\em arXiv preprint arXiv:1807.03748}, 2018.

\bibitem{DBLP:conf/interspeech/SchneiderBCA19}
Steffen Schneider, Alexei Baevski, Ronan Collobert, and Michael Auli,
\newblock ``wav2vec: Unsupervised pre-training for speech recognition,''
\newblock in {\em Interspeech 2019}, Gernot Kubin and Zdravko Kacic, Eds. 2019,
  pp. 3465--3469, {ISCA}.

\bibitem{DBLP:conf/iclr/BaevskiSA20}
Alexei Baevski, Steffen Schneider, and Michael Auli,
\newblock ``vq-wav2vec: Self-supervised learning of discrete speech
  representations,''
\newblock in {\em {ICLR} 2020}. 2020, OpenReview.net.

\bibitem{DBLP:conf/nips/BaevskiZMA20}
Alexei Baevski, Yuhao Zhou, Abdelrahman Mohamed, and Michael Auli,
\newblock ``wav2vec 2.0: {A} framework for self-supervised learning of speech
  representations,''
\newblock in {\em NeurIPS 2020}, Hugo Larochelle, Marc'Aurelio Ranzato, Raia
  Hadsell, Maria{-}Florina Balcan, and Hsuan{-}Tien Lin, Eds., 2020.

\bibitem{xu2021self}
Qiantong Xu, Alexei Baevski, Tatiana Likhomanenko, Paden Tomasello, Alexis
  Conneau, Ronan Collobert, Gabriel Synnaeve, and Michael Auli,
\newblock ``Self-training and pre-training are complementary for speech
  recognition,''
\newblock in {\em ICASSP 2021-2021 IEEE International Conference on Acoustics,
  Speech and Signal Processing (ICASSP)}. IEEE, 2021, pp. 3030--3034.

\bibitem{zhang2020pushing}
Yu~Zhang, James Qin, Daniel~S Park, Wei Han, Chung-Cheng Chiu, Ruoming Pang,
  Quoc~V Le, and Yonghui Wu,
\newblock ``Pushing the limits of semi-supervised learning for automatic speech
  recognition,''
\newblock {\em arXiv preprint arXiv:2010.10504}, 2020.

\bibitem{baevski2021unsupervised}
Alexei Baevski, Wei-Ning Hsu, Alexis Conneau, and Michael Auli,
\newblock ``Unsupervised speech recognition,''
\newblock {\em arXiv preprint arXiv:2105.11084}, 2021.

\bibitem{DBLP:conf/icassp/RiviereJMD20}
Morgane Rivi{\`{e}}re, Armand Joulin, Pierre{-}Emmanuel Mazar{\'{e}}, and
  Emmanuel Dupoux,
\newblock ``Unsupervised pretraining transfers well across languages,''
\newblock in {\em {ICASSP} 2020}. 2020, pp. 7414--7418, {IEEE}.

\bibitem{DBLP:journals/corr/abs-2006-13979}
Alexis Conneau, Alexei Baevski, Ronan Collobert, Abdelrahman Mohamed, and
  Michael Auli,
\newblock ``Unsupervised cross-lingual representation learning for speech
  recognition,''
\newblock {\em CoRR}, vol. abs/2006.13979, 2020.

\bibitem{wang2021unispeech}
Chengyi Wang, Yu~Wu, Yao Qian, Kenichi Kumatani, Shujie Liu, Furu Wei, Michael
  Zeng, and Xuedong Huang,
\newblock ``Unispeech: Unified speech representation learning with labeled and
  unlabeled data,''
\newblock {\em arXiv preprint arXiv:2101.07597}, 2021.

\bibitem{chan2021speechstew}
William Chan, Daniel Park, Chris Lee, Yu~Zhang, Quoc Le, and Mohammad Norouzi,
\newblock ``Speechstew: Simply mix all available speech recognition data to
  train one large neural network,''
\newblock {\em arXiv preprint arXiv:2104.02133}, 2021.

\bibitem{sainath2020streaming}
Tara~N Sainath, Yanzhang He, Bo~Li, et~al.,
\newblock ``A streaming on-device end-to-end model surpassing server-side
  conventional model quality and latency,''
\newblock in {\em Proc. ICASSP}, 2020, pp. 6059--6063.

\bibitem{Li2020Developing}
Jinyu Li, , Rui Zhao, Zhong Meng, et~al.,
\newblock ``Developing {RNN-T} models surpassing high-performance hybrid models
  with customization capability,''
\newblock in {\em Proc. Interspeech}, 2020.

\bibitem{yeh2019transformer}
Ching-Feng Yeh, Jay Mahadeokar, Kaustubh Kalgaonkar, Yongqiang Wang, Duc Le,
  Mahaveer Jain, Kjell Schubert, Christian Fuegen, and Michael~L Seltzer,
\newblock ``Transformer-transducer: End-to-end speech recognition with
  self-attention,''
\newblock {\em arXiv preprint arXiv:1910.12977}, 2019.

\bibitem{zhang2020transformer}
Qian Zhang, Han Lu, Hasim Sak, Anshuman Tripathi, Erik McDermott, Stephen Koo,
  and Shankar Kumar,
\newblock ``Transformer transducer: A streamable speech recognition model with
  transformer encoders and rnn-t loss,''
\newblock in {\em ICASSP 2020-2020 IEEE International Conference on Acoustics,
  Speech and Signal Processing (ICASSP)}. IEEE, 2020, pp. 7829--7833.

\bibitem{chen2021developing}
Xie Chen, Yu~Wu, Zhenghao Wang, Shujie Liu, and Jinyu Li,
\newblock ``Developing real-time streaming transformer transducer for speech
  recognition on large-scale dataset,''
\newblock in {\em Proc. ICASSP}, 2021.

\bibitem{DBLP:conf/interspeech/ChungHTG19}
Yu{-}An Chung, Wei{-}Ning Hsu, Hao Tang, and James~R. Glass,
\newblock ``An unsupervised autoregressive model for speech representation
  learning,''
\newblock in {\em Interspeech 2019}, Gernot Kubin and Zdravko Kacic, Eds. 2019,
  pp. 146--150, {ISCA}.

\bibitem{DBLP:conf/icassp/LiuYCHL20}
Andy~T. Liu, Shu{-}Wen Yang, Po{-}Han Chi, Po{-}chun Hsu, and Hung{-}yi Lee,
\newblock ``Mockingjay: Unsupervised speech representation learning with deep
  bidirectional transformer encoders,''
\newblock in {\em {ICASSP} 2020}. 2020, pp. 6419--6423, {IEEE}.

\bibitem{DBLP:conf/icassp/LingLSK20}
Shaoshi Ling, Yuzong Liu, Julian Salazar, and Katrin Kirchhoff,
\newblock ``Deep contextualized acoustic representations for semi-supervised
  speech recognition,''
\newblock in {\em {ICASSP} 2020}. 2020, pp. 6429--6433, {IEEE}.

\bibitem{ling2020bertphone}
Shaoshi Ling, Julian Salazar, Yuzong Liu, Katrin Kirchhoff, and AWS Amazon,
\newblock ``Bertphone: Phonetically-aware encoder representations for
  utterance-level speaker and language recognition,''
\newblock in {\em Proc. Odyssey 2020 The Speaker and Language Recognition
  Workshop}, 2020, pp. 9--16.

\bibitem{DBLP:conf/emnlp/KawakamiWDBO20}
Kazuya Kawakami, Luyu Wang, Chris Dyer, Phil Blunsom, and A{\"{a}}ron van~den
  Oord,
\newblock ``Learning robust and multilingual speech representations,''
\newblock in {\em {EMNLP} 2020}, Trevor Cohn, Yulan He, and Yang Liu, Eds.
  2020, pp. 1182--1192, Association for Computational Linguistics.

\bibitem{vaswani2017attention}
Ashish Vaswani, Noam Shazeer, Niki Parmar, Jakob Uszkoreit, Llion Jones,
  Aidan~N Gomez, Lukasz Kaiser, and Illia Polosukhin,
\newblock ``Attention is all you need,''
\newblock {\em arXiv preprint arXiv:1706.03762}, 2017.

\bibitem{DBLP:conf/iclr/BaevskiA19}
Alexei Baevski and Michael Auli,
\newblock ``Adaptive input representations for neural language modeling,''
\newblock in {\em 7th International Conference on Learning Representations,
  {ICLR} 2019, New Orleans, LA, USA, May 6-9, 2019}. 2019, OpenReview.net.

\bibitem{DBLP:conf/icml/XiongYHZZXZLWL20}
Ruibin Xiong, Yunchang Yang, Di~He, Kai Zheng, Shuxin Zheng, Chen Xing,
  Huishuai Zhang, Yanyan Lan, Liwei Wang, and Tie{-}Yan Liu,
\newblock ``On layer normalization in the transformer architecture,''
\newblock in {\em Proceedings of the 37th International Conference on Machine
  Learning, {ICML} 2020, 13-18 July 2020, Virtual Event}. 2020, vol. 119 of
  {\em Proceedings of Machine Learning Research}, pp. 10524--10533, {PMLR}.

\bibitem{DBLP:conf/naacl/ShawUV18}
Peter Shaw, Jakob Uszkoreit, and Ashish Vaswani,
\newblock ``Self-attention with relative position representations,''
\newblock in {\em Proceedings of the 2018 Conference of the North American
  Chapter of the Association for Computational Linguistics: Human Language
  Technologies, NAACL-HLT, New Orleans, Louisiana, USA, June 1-6, 2018, Volume
  2 (Short Papers)}, Marilyn~A. Walker, Heng Ji, and Amanda Stent, Eds. 2018,
  pp. 464--468, Association for Computational Linguistics.

\bibitem{graves2006connectionist}
Alex Graves, Santiago Fern{\'a}ndez, Faustino Gomez, and J{\"u}rgen
  Schmidhuber,
\newblock ``Connectionist temporal classification: labelling unsegmented
  sequence data with recurrent neural networks,''
\newblock in {\em Proceedings of the 23rd international conference on Machine
  learning}, 2006, pp. 369--376.

\bibitem{Graves-RNNSeqTransduction}
Alex Graves,
\newblock ``Sequence transduction with recurrent neural networks,''
\newblock in {\em http://arxiv.org/abs/1211.3711}, 2012.

\bibitem{li2012improving}
Jinyu Li, Dong Yu, Jui-Ting Huang, and Yifan Gong,
\newblock ``Improving wideband speech recognition using mixed-bandwidth
  training data in {CD-DNN-HMM},''
\newblock in {\em Proc. SLT}. IEEE, 2012, pp. 131--136.

\end{thebibliography}

\end{document}